\documentstyle[12pt]{article}
\def\fun#1#2{\lower3.6pt\vbox{\baselineskip0pt\lineskip.9pt
  \ialign{$\mathsurround=0pt#1\hfil##\hfil$\crcr#2\crcr\sim\crcr}}}

\title{\bf
QED Radiative Corrections to the Pionium Life Time.
}

\author{E.A.~Kuraev}

\date{}

\begin{document}

\maketitle

\vskip 0.3cm

\begin{center}
{\it Joint Institute for Nuclear Research,
Dubna, Moscow region, 141980, Russia }
\end{center}

\vskip 1.5cm
\begin{abstract}
The lowest order QED radiative corrections to the cross section of the
recharged process of transition
of two charged pions to two neutral ones and to the pionium life time
are calculated in frames of scalar QED. It is argued that the ultraviolet
cut--off of the loop momentum is to be chosen of order of $\rho$ meson
mass. This fact permits to perform the calculation in frames of Effective
Chiral Lagrangian theory with vector-meson dominance. The
Coulomb factor corresponding to interaction in the initial state, shown,
is to be removed to avoid the double counting. Resulting value of the
radiative correction to the pionium life time is $-0.25\%$.
\end{abstract}

\section{Introduction}

The precise measurement of the pionium life time~[1] provide a crucial
experiment to verify the (different) predictions of the Chiral
Perturbation Theory
(CHPT)~concerning pion--pion interaction~\cite{r2a,r2b}. Considering the
conversion of two charged pions to two
neutral due to strong interactions as the main mechanism\footnote{The
two quantum annihilation mechanism may be separated experimentally as
well as it corresponds to photon energy (in the pionium center-of-mass
frame) equal to pion mass. The background from four photon annihilation
channel is of order $(\alpha/\pi)^2$, i.e. negligible.}
which determine
the pionium lifetime $\tau$ the formula for it was derived many years
ago~\cite{r3}:
\begin{equation}
\tau^{-1}_0=\frac{16\pi\sqrt{2}}{9}
(1-\frac{m_{\pi_0}}{m_{\pi_+}})^{\frac{1}{2}} \nonumber \\
(a_0-a_2)^2|\Psi_{n,0}(0)|^2[1+\frac{2}{9}m_{\pi_+}
(m_{\pi_+}-m_{\pi_0})(2a_2+a_0)^2]^{-1}.
\end{equation}
Right-hand side of (1) have a factorized form. One factor $|\Psi_{n,0}(0)|^2$
have a pure Coulomb interaction origin and describe the probability
to form the pionium $s$-wave state. Another, $(a_0-a_2)^2$, describes the
low-energy pions-conversion process and the quantities $a_0,a_2$ are
the scattering lengths in the states with the isotopic spin $0,2$.
Remaining factors have a kinematical origin.

The problem of calculation of corrections to $\tau^{-1}$ is rather
subtle one. First, the isotopic-spin classification is violated by
electromagnetic interactions, the electromagnetic interactions themselves
provide the existence of pionium atom, and, finally, the influence of the
strong interactions on the value of the wave function at zero distance
is to be taken into account. So I propose such form of the corrected
pionium lifetime:
\begin{equation}
\tau^{-1}=\tau^{-1}_0(1+\delta_{\psi})(1+\delta)(1+\delta_a).
\end{equation}

The quantities $\tau^{-1}_0$ is given in (1) and corresponds to the case
when all corrections are switched off;
$\delta_{\psi}$ include the corrections arising
from modifications of Bethe--Salpiter equation for $\Psi(r)$ due to strong
interactions~\cite{r4}, the quantity $\delta_a$ include the higher orders
of CHPT contributions~\cite{r2a}.
Main attention here will be paid to calculation of the pure QED
correction (RC), $\delta$, in the lowest order of perturbation theory (PT).

Paper is organized as follows.In part 1 the recharged
process
\begin{equation}
\pi^+(p_2)+\pi^-(p_1) \rightarrow  \pi^0(q_1)+\pi^0(q_2)
\end{equation}
for the case of free pions moving with the small relative
velocity $v\approx 2\beta$, is considered using the approximation of the
point--like pions interacting with the electromagnetic field.
Calculating the virtual
corrections we meet as usually the infrared, ultraviolet divergences,
and the known Coulomb factor which describe the Coulomb interaction
of the slow moving charged particles. It is shown that the last factor
corresponds to small values of the loop momenta and is to be omitted
when considering the Coulomb interaction corrected cross section
of recharged process (3) (the similar problem arose in calculation
of RC to parapositronium width calculation~\cite{r5}).
The infrared singularities are removed by the usual way when
taking into account the emission of real soft photons. Some arguments
are given to choice the ultraviolet cut-off parameter to be of order
of $\rho$-meson mass, $\Lambda=m_{\rho}$.
Within the same approach we also consider the realisitic case of
bounded pions as components of pionium atom.
In the second part we perform the calculations in frame of effective
chiral lagrangian with vector meson dominance~\cite{bpr}.

1.Consider first the RC to the cross section of the conversion process
in frames of scalar electrodynamics when pions directly interact with
the electromagnetic field.
We will suppose the pions be real $p_i^2=m_+^2=m^2$, $q_i^2=m_0^2$, $i=1,2$
and that they have a small relative velocity $v=2\beta=2\sqrt{1-
\frac{m^2}{\varepsilon^2}}$, where $\varepsilon$ is the energy of one
of pions in the cms. The lowest order virtual correction in the case of
point--like pion are determined by quantity $\delta_{virt}$:
\begin{eqnarray}
\delta_{virt}&=&\frac{2ReM_0M^{(1)*}}{|M_0|^2}=\frac{\alpha}{2\pi}(2a+b),
\nonumber \\
a&=&\ln\frac{\Lambda^2}{m^2}+\ln\frac{m^2}{\lambda^2}-\frac{3}{4},
\\ \nonumber
b&=&\int dk\frac{(2p_1-k)(-2p_2-k)}{(k^2-\lambda^2)(k^2-2p_1k)(k^2+2p_2k)},
\\ \nonumber
|k^2| &<& \Lambda^2,\qquad dk=\frac{d^4k}{i\pi^2}.
\end{eqnarray}
In this equation $\Lambda,\lambda $ are the ultraviolet, infrared momentum
cut-off's,$m$ is the renormalized pion mass. The quantity $a$ is related
with the pion wave function renormalization:
\begin{equation}
G_{\pi}(k)=\frac{i}{k^2-m_{bare}^2-\Sigma(k)}\rightarrow \frac{iZ^f}{k^2-m^2},
\quad Z^f=1+\frac{\alpha}{2\pi}a.
\end{equation}

The standard calculation of the loop integral give
\begin{eqnarray}
&& b=\int_0^1 dx [\ln\frac{\Lambda^2}{p_x^2}-1+\frac{2m^2(1+\beta^2)}{(1-\beta^2)p_x^2}
\ln\frac{p_x^2}{\lambda^2} - \frac{4m^2\beta^2}{p_x^2(1-\beta^2)}],
\\ \nonumber
&& p_x^2=\frac{m^2}{1-\beta^2}[(1-2x)^2-\beta^2]-i0.
\end{eqnarray}
Using the I.~Harris and L.~Brown~\cite{r5} result
\begin{equation}
\mbox{Re}\; \int_0^1\frac{dx m^2}{p_x^2}ln\frac{p_x^2}{\lambda^2}=-(1-\frac{2}{3}\beta^2)(2+\ln\frac{m^2}{\lambda^2})
-\frac{2}{9}\beta^2+\frac{\pi^2(1-\beta^2)}{2\beta}+0(\beta^3),
\end{equation}
we obtain for contribution of loop integrals:
\begin{eqnarray}
\delta_{virt}&=&\frac{\alpha}{2\pi}
[3 \ln\frac{\Lambda^2}{m^2}-\frac{8}{3}\beta^2
\ln\frac{m^2}{\lambda^2} \nonumber \\
&-&\frac{9}{2}-\frac{34}{9}\beta^2+\frac{\pi^2}{\beta}(1+\beta^2)]
+O(\frac{\alpha}{\pi}\beta^3).
\end{eqnarray}

The contribution from emission of real photon (we may use here the
soft photon emission approximation) gives:
\begin{eqnarray}
\delta_{soft}&=&-\frac{\alpha}{4\pi^3}
\int\frac{d^3k}{\sqrt{\lambda^2+k^2}}(\frac{p_1}{p_1k}
-\frac{p_2}{p_2k})^2,|k|<\delta \varepsilon=m-m_0,  \nonumber \\
\delta_{soft}&=&\frac{\alpha}{\pi}\beta^2
(\frac{8}{3}\ln\frac{2\delta \varepsilon}{\lambda}-\frac{20}{9})
+O(\frac{\alpha}{\pi}\beta^3).
\end{eqnarray}
In the total sum $\delta_{virt}+\delta_{soft}$ the dependence on the
photon mass $\lambda$ disappears. Keeping in mind that the effective
velocities in pionium atom are of order of fine structure constant (in
units of light velocity) we may neglect the contributions of order
$\frac{\alpha}{\pi}\beta^2$.

Now we note that the term $\frac{\pi^2}{\beta}(1+\beta^2)$ in $\delta_{virt}$
(8) is the lowest order of expansion of the known Coulomb factor
$J(x)$ which is to be included in matrix element module squared with
charged particles in the initial state. This factor
describe the Coulomb interaction of charged
particles with the small relative velocity $v$. In the case of equal
mass and opposite charges it have a form ($ x=\frac{\alpha}{v},
v \approx 2\beta$):
\begin{equation}
J(x)=\frac{2\pi x}{1-\exp\left\{-2\pi x\right\}}
=1+\frac{\pi\alpha}{v}.
\end{equation}
It arises from the region of small energies and 3--momenta of virtual
loop photons $|k^0|\simeq m\alpha^2,|\vec {k}|\simeq m\alpha$.

Now we argue that the natural choice of the ultraviolet cut--off parameter $\Lambda$
is the typical vector meson mass,$\Lambda=m_{\rho}$.Really,at large values
of loop momenta $|k|$ pion is to be considered as a quark-antiquark system.
The conversion process a this level may be interpreted as an exchange by
$u,d$ or $\bar{u},\bar{d}$ quarks accompanied by (multi)gluon exchange.
By simple power counting one may see that the virtual photon loop intergals
are convergent. The natural scale of the loop momenta which provide its
main contribution of order of typical hadron's mass,$|k|\simeq m_{\rho}$.
This quantity play the role of the ultraviolet cut-off parameter, $\Lambda$.
Really,the theoretical uncertainty of order $\alpha/\pi$ in the RC will
appear in such an approach as well as the confinement mechanism is purely
investigated.
Note that the choice $\Lambda=m_{\rho}$ crucially differs from the $\Lambda=
M_Z$ which is natural for the case of semileptonic decays of pions~\cite{r6,r7}.

For the coulomb-corrected cross section of conversion process (6)
$\sigma=J(x)\sigma_c$ we obtain:
\begin{equation}
\sigma_c=\sigma_0(1+\delta_{QED}^f),\delta_{QED}^f=\frac{\alpha}{\pi}[3\ln\frac{\Lambda}{m}-
\frac{9}{4}+0(1)],\Lambda\approx {m_{\rho}}.
\end{equation}
and $\sigma_0$ is the cross section without QED corrections. Main source
of errors here due to uncertainty in the quantity $\Lambda$.

More exact results may be obtained in concrete models, such as the quark
model. We will consider below the Effective Chiral Lagrangian with vector
meson dominance, the particle analog of the quark model for this aim.

In the same frames of scalar QED we consider now the case of bound pions,
$p_1^2-m^2=p_2^2-m^2=-\Delta$, where $\Delta\approx {m^2\alpha^2}$,
i.e. the correction to pionium life time.

Again the virtual corrections will have a form as in previous case with
the replacements
\begin{equation}
a\rightarrow a'=\ln\frac{\Lambda^2}{m^2}+2\ln\frac{m^2}{\Delta}-\frac{3}{4};
b\rightarrow b'=\ln\frac{\Lambda^2}{m^2}+1+
\int\frac{dx}{p_x^2}ln\frac{p_x^2}{\Delta}+0(\beta^2).
\end{equation}
In this case the emission of the real photons is forbidden. The Coulomb factor
is to be removed to avoid the double counting as well as it is taken into account
in the factor $\|\Psi(o)\|^2$ in (1). The resulting expression for the
correction $\delta$ (2) have a form:
\begin{equation}
\delta_{QED}=\frac{\alpha}{\pi}(3\ln\frac{\Lambda}{m}-\frac{25}{4}+0(1)),\Lambda=m_{\rho}.
\end{equation}

2.In frames of Effective Chiral Lagrangian with the vector meson dominance~\cite{bpr}:
\begin{equation}
L=\frac{1}{2}[m_{\rho}^2\rho_{\mu}^2+(d_{\mu}\pi^+)^2+(d_{\mu}\pi^-)^2]-
\frac{e}{g}m_{\rho}^2A_{\mu}\rho_{\mu}-ig\rho_{\mu}(\pi^+(d_{\mu}\pi_-)-
\pi^-(d_{\mu}\pi_+))+...
\end{equation}
with irrelevant terms omitted, the direct interaction of pions to
electromagnetic field absent.

The similar calculations\footnote{Technically it results as a factor
$(m_{\rho}^2/(k^2-m_{\rho}^2))^2$ in the
integrand for $a$ and the similar change in calculation $Z_{\pi}$.Loop
momentum integrals become ultraviolet convergent.}give the values of
$Z_{\pi}$ for the cases of free pions and the bound ones:
\begin{eqnarray}
Z_{\pi}^f & = & 1+\frac{\alpha}{2\pi}[\ln\frac{m_{\rho}^2}{m^2}+\ln\frac{m^2}{\lambda^2}
-\frac{7}{4}+\frac{m^2}{m_{\rho}^2}(\ln\frac{m_{\rho}^2}{m^2}-\frac{7}{6})+0((\frac{m}{m_{\rho}})^4)],
\nonumber\\
Z_{\pi} & = & 1+\frac{\alpha}{2\pi}[\ln\frac{m_{\rho}^2}{m^2}+2\ln\frac{m^2}{\Delta}
-\frac{7}{4}+\frac{m^2}{m_{\rho}^2}(\ln\frac{m_{\rho}^2}{m^2}-\frac{7}{6})+0((\frac{m}{m_{\rho}})^4)].
\end{eqnarray}
The final expressions for $\delta^f,\delta$ have a form:
\begin{eqnarray}
\delta^f & = & \frac{\alpha}{2\pi}[3\ln\frac{m_{\rho}^2}{m^2}-\frac{15}{2}-
\frac{m^2}{m_{\rho}^2}(\frac{16}{3}\ln\frac{m_{\rho}^2}{m^2}+\frac{98}{9})+0((\frac{m}{m_{\rho}})^4)],
\nonumber\\
\delta & = & \frac{\alpha}{2\pi}[3\ln\frac{m_{\rho}^2}{m^2}-\frac{23}{2}-
\frac{m^2}{m_{\rho}^2}(\frac{16}{3}\ln\frac{m_{\rho}^2}{m^2}+\frac{98}{9})+0((\frac{m}{m_{\rho}})^4)].
\end{eqnarray}
Numerically these quantities close to the case of scalar QED considered above.
The contributions from Feynman graphs containing $\rho\pi\omega$ vertex
turns out to be small (do not exceed $0.1\%$).
We do not consider here the contributions arising from excited mesons
in the intermediate state. It is, presumably also small (do not exceed
$0.1\%$) due to smallness of their coupling to the pions. This question
remains open.

The author is indebted to L.L.~Nemenov who turns my attention
to this problem and to J.~Gasser, A.~Smilga, M.~Vysotsky,
V.~Novikov, M.~Volkov, S.~Gerasimov, A.V.~Tarasov and
A.~Arbuzov for discussions.

\end{document}